\documentclass[twocolumn,preprintnumbers,amsmath,amssymb,superscriptaddress]{revtex4}

\usepackage{graphicx}
\usepackage{dcolumn}
\usepackage{bm}
\usepackage{soul}
\usepackage{color}
\usepackage{epstopdf}
\usepackage[version=3]{mhchem}
\usepackage{lipsum}
\usepackage[outercaption]{sidecap}
\usepackage{floatrow}
\begin{document}

\title{Effects of Cooling Rate on Structural Relaxation in Amorphous Drugs: Elastically Collective Nonlinear Langevin Equation Theory and Machine Learning Study}

\author{Anh D. Phan}
\affiliation{Faculty of Materials Science and Engineering, Phenikaa Institute for Advanced Study, Phenikaa University, Hanoi 100000, Vietnam}
\email{anh.phanduc@phenikaa-uni.edu.vn}
\affiliation{Faculty of Information Technology, Artificial Intelligence Laboratory, Phenikaa University, Hanoi 100000, Vietnam}
\affiliation{Department of Nanotechnology for Sustainable Energy, School of Science and Technology, Kwansei Gakuin University, Sanda, Hyogo 669-1337, Japan}
\author{Katsunori Wakabayashi}
\affiliation{Department of Nanotechnology for Sustainable Energy, School of Science and Technology, Kwansei Gakuin University, Sanda, Hyogo 669-1337, Japan}
\author{Marian Paluch}
\affiliation{Institute of Physics, University of Silesia, SMCEBI, 75 Pułku Piechoty 1a, 41-500 Chorzow, Poland}
\author{Vu D. Lam}
\affiliation{Institute of Materials Science, Vietnam Academy of Science and Technology, Hanoi, Viet Nam}
\affiliation{Graduate University of Science and Technology, Vietnam Academy of Science and Technology, Hanoi, Viet Nam}
\date{\today}

\date{\today}

\begin{abstract}
Theoretical approaches are formulated to investigate the molecular mobility under various cooling rates of amorphous drugs. We describe the structural relaxation of a tagged molecule as a coupled process of cage-scale dynamics and collective molecular rearrangement beyond the first coordination shell. The coupling between local and non-local dynamics behaves distinctly in different substances. Theoretical calculations for the structural relaxation time, glass transition temperature, and dynamic fragility are carried out over twenty-two amorphous drugs and polymers. Numerical results have a quantitatively good accordance with experimental data and the extracted physical quantities using the Vogel-Fulcher-Tammann fit function and machine learning. The machine learning method reveals the linear relation between the glass transition temperature and the melting point, which is a key factor for pharmaceutical solubility. Our predictive approaches are reliable tools for developing drug formulation. 
\end{abstract}

\maketitle

\section{Introduction}
Amorphous drugs have attracted much attention \cite{49,50,51,52,53} owing to large solubility and enhanced bioavailability compared to the crystalline counterparts. The disordered structure of amorphous pharmaceutical products is formed by the rapid cooling of the molten material. The molecular mobility of an amorphous material is characterized by structural (alpha) relaxation time, $\tau_\alpha$. Since the structural relaxation process originates from liquid structure reorganization, $\tau_\alpha$ is temperature-dependent and significantly slowed down at low temperatures. Below the glass transition temperature $T_g$, which is defined by $\tau_\alpha(T_g) = 100$ s, the drug stays in a disordered state for long time, which is larger than the experimental observation time scale. However, the pharmaceutical can possibly be recrystallized during manufacturing or storage processes \cite{49,50}. It turns out that the physical stability of many amorphous systems is relatively poor. Comprehensive understanding of glassy states and molecular mobility of amorphous drugs is crucial to formulate pharmaceutical products having desired properties \cite{49,50} and understand fundamentals of glassy state physics.

The relaxation processes of amorphous materials can be experimentally measured using broadband dielectric spectroscopy (BDS) and differential scanning calorimetry (DSC) \cite{49,50}. BDS technique determines the structural relaxation time corresponding to thermal variation. DSC can measure the glass transition temperature and analyze phase separation in experimental samples at different cooling rates. The relevant timescale of molecular motions measured by BDS spans from picosecond above melting temperature to hundreds seconds in vicinity of the glass transition temperature. This technique can be used to investigate both structural (primary or long-time) and transient (secondary or short-time) relaxation processes. 

Recently, the Elastically Collective Nonlinear Langevin Equation (ECNLE) theory has been developed to understand structural relaxation time of amorphous systems \cite{2,6,7,10,8,42}, in which an amorphous material is modeled as a fluid of molecular particles. The ECNLE theory considers a single molecular motion affected by the nearest-neighbor interactions and cooperative motions of molecules outside a particle cage formed by the neighboring molecules \cite{2,6,7,10,8}. This physical picture leads to a local barrier of a dynamic free energy and a collective barrier for each molecule caused by its nearest-neighbor interactions and effects of cooperative molecular rearrangements, respectively. These barriers are density-dependent and mutually correlated. 

Plugging the two barriers calculated using the ECNLE theory into Kramer's theory gives the structural relaxation time \cite{2,6,7,10,8,42}. To find the temperature dependence of alpha relaxation time, one proposed an analytical conversion (thermal mapping) from averaged particle density to temperature based on experimental dimensionless compressibility data associated with equation of state \cite{2,6,7,10,8}. The theoretical calculations have also provided quantitative good predictions for the glass transition temperature and dynamic fragility of colloidal systems \cite{10,7}, supercooled liquids \cite{6,10}, and polymer melts \cite{2,8}. However, the experimental equation-of-state data needed for the original thermal mapping has rarely measured in amorphous drugs. Thus, we have recently proposed another density-to-temperature conversion using the thermal expansion and experimental glass transition temperature values \cite{42}. Our approach has successfully described temperature-dependent molecular dynamics in one- and multi-component amorphous drugs. 

Most ECNLE calculations have assumed a universal correlation between local and collective barrier for all substances when inserting into Kramer's theory. The assumption simplifies roles of chemical and biological complexities on the glass transition. Consequently, the quantitative agreement between theoretical calculations and experiments is imperfect. Moreover, since the dynamic fragility of a material depends strongly on the form of $\tau_\alpha(T)$, a small theory-experiment deviation in $\tau_\alpha(T)$ leads to an inaccurate prediction of the fragility.

In this work, we introduce an adjustable parameter characterizing for a non-universal local-collective correlation in pharmaceuticals. The new version of the ECNLE theory accurately and simultaneously predicts the glass transition temperature and dynamic fragility of amorphous drugs. We employ machine learning technique to reveal a undiscovered relation between melting point and glass transition. We also predict the glass transition temperature based on BDS data. Our ECNLE numerical results are quantitatively compared to experimental data and machine learning calculations.

Theoretical understanding of how glassy dynamics of amorphous materials varies with the cooling rate is limited. Previous theoretical studies \cite{44,12,13} have given phenological/qualitative descriptions for the $T_g$ change as a function of cooling rate. Based on the theoretical development of this paper, we propose, for the first time, an analysis to provide quantitative determinations for this phenomenon and insightful discussions for experiments. 

\section{Structural relaxation time of amorphous drugs}
Amorphous drugs/materials are theoretically described as a fluid of disconnected spherical rigid particles, which interact with each other via hard-sphere interaction. The particle diameter is $d$ and the number density of particles is $\rho$. A particle is assumingly acted by three forces including (1) the caging force caused by the surrounding fluid, -$\partial F_{dyn}(r)/\partial r$, (2) the thermal white noise, $\delta f$, (3) the friction force, $-\zeta_s(\partial r/\partial t)$, here $\zeta_s$ is a short-time friction constant and $r\equiv r(t)$ is the displacement of the particle. The key quantity is $F_{dyn}(r)$ known as the dynamic free energy of the tagged particle due to its nearest neighbors \cite{2,3,4,6}. Based on an overdamped equation-of-motion for the scalar displacement of a tagged particle, we have
\begin{eqnarray}
-\zeta_s\frac{\partial r}{\partial t} -\frac{\partial F_{dyn}(r)}{\partial r} + \delta f = 0.
\label{eq:1}
\end{eqnarray}
The thermal noise force satisfies $\left<\delta f(0)\delta f(t) \right>=2k_BT\zeta_s\delta(t)$, where $k_B$ is the Boltzmann constant and $T$ is temperature. According to ECNLE theory, the dynamic free energy is \cite{2,3,4,6}
\begin{eqnarray}
\frac{F_{dyn}(r)}{k_BT} &=& \int_0^{\infty} dq\frac{ q^2d^3 \left[S(q)-1\right]^2}{12\pi\Phi\left[1+S(q)\right]}\exp\left[-\frac{q^2r^2(S(q)+1)}{6S(q)}\right]
\nonumber\\ &-&3\ln\frac{r}{d},
\label{eq:2}
\end{eqnarray}
where $\Phi = \rho\pi d^3/6$ is the volume fraction, $q$ is the wavevector, and the static structure factor $S(q)$ is calculated using Percus-Yevick (PY) integral equation theory \cite{1} for a hard-sphere fluid. The PY theory expresses $S(q)$ via the direct correlation function $C(q)=\left[S(q)-1 \right]/\rho S(q)$. While the fourier transform of $C(q)$ or the real-space direct correlation function is \cite{1}
\begin{eqnarray}
C(r) &=& -\frac{(1+2\Phi)^2}{(1-\Phi)^4} + \frac{6\Phi(1+\Phi/2)^2}{(1-\Phi)^4}\frac{r}{d}\nonumber\\
&-&\frac{\Phi(1+2\Phi)^2}{2(1-\Phi)^4}\left(\frac{r}{d}\right)^3 \quad \mbox{for} \quad r \leq d \\
C(r) &=& 0 \quad \mbox{for} \quad r > d.
\end{eqnarray}
Recall that the ECNLE theory ignores effects of rotational motions and only consider translational motions, which are angularly-averaged. The first term of Eq. (\ref{eq:2}), which depends strongly on the fluid structure and density, is the dynamic trapping potential and favors the particle localization. While the second term independent of the system structure represents the ideal fluid state.

\begin{figure}[htp]
\center
\includegraphics[width=8.5cm]{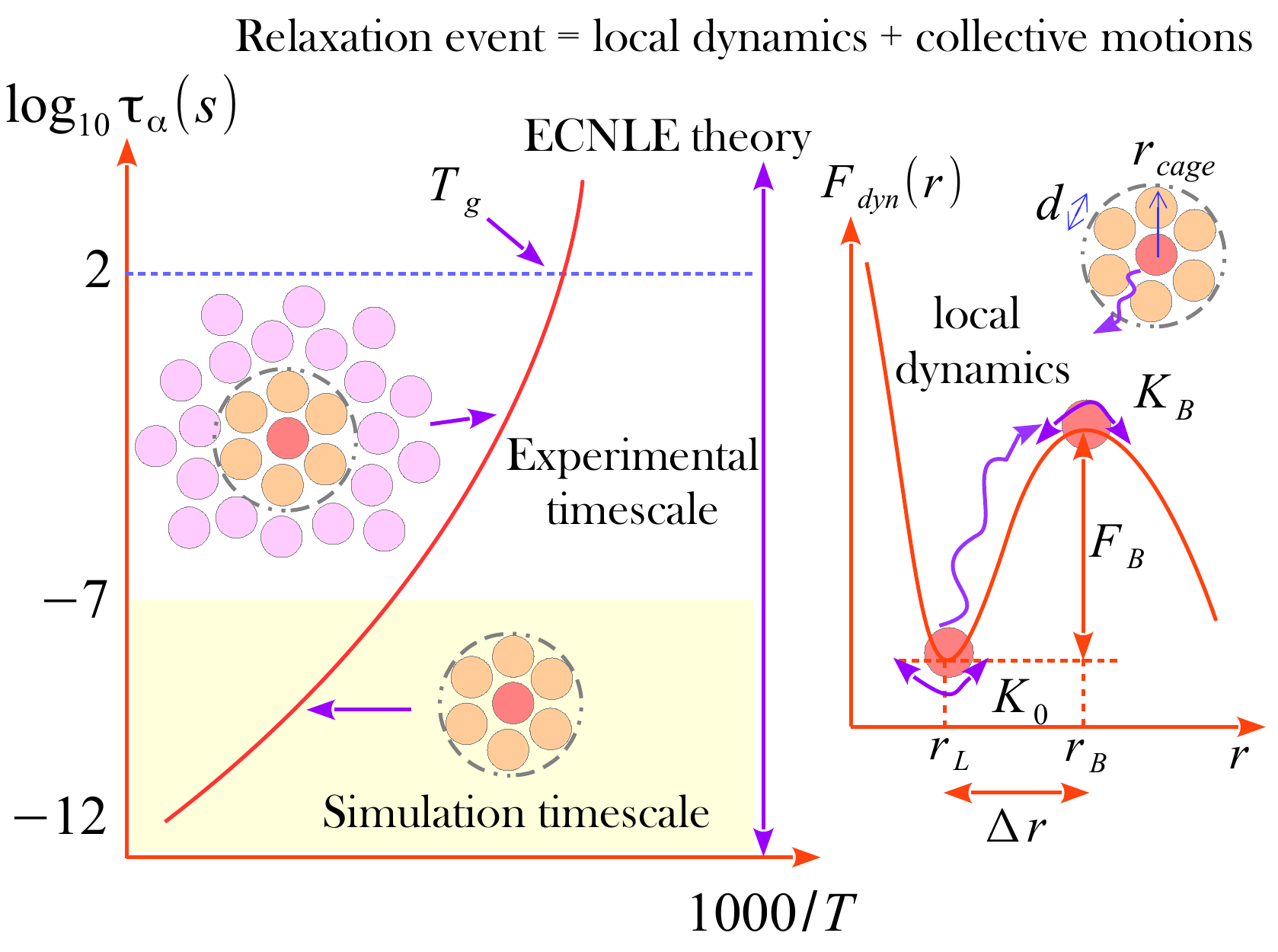}
\caption{\label{fig:1}(Color online) A general form of temperature dependence of structural relaxation time described by a coupled process of local cage-scale dynamics and collective motions. The dynamic free energy profile of a tagged particle indicates key length scales and a barrier in the local dynamics.}
\end{figure}

Figure \ref{fig:1} shows an illustration of the dynamic free energy as a function of $r$ and key physical quantities in the local caging constraint. In dilute suspension ($\Phi \leq 0.43$) \cite{3,4,1}, $F_{dyn}(r)$ monotonically decreases with an increase of $r$ and particles move without constraint. When $\Phi > 0.43$, the tagged particle is dynamically arrested within a particle cage formed by its neighbors and one observes an emergence of a free-energy barrier. The particle cage radius, $r_{cage}$, as depicted in Fig. \ref{fig:1} is determined as the first minimum position in the radial distribution function, $g(r)$. Since $S(q)$ and $g(r)$ are a Fourier-transform pair, one has $g(r)-1=\frac{1}{2\pi^2\rho r} \int_0^{\infty}\left[S(q)-1\right]q\sin(qr)dq$. Thus, the radius of the particle cage is about $1.3-1.5d$. When the dynamic free energy reaches the local minimum and maximum, one obtains the localization length ($r_L$) and the barrier position ($r_B$). The energy difference between these two positions is the local hopping barrier $F_B=F_{dyn}(r_B)-F_{dyn}(r_L)$. The jump distance from the localized position to the barrier position is defined as $\Delta r =r_B-r_L$. $K_0 = \left|\partial^2 F_{dyn}(r)/\partial r^2\right|_{r=r_L}$ and $K_B$=$\left|\partial^2 F_{dyn}(r)/\partial r^2\right|_{r=r_B}$ are absolute curvatures at the localization length and barrier position.

The rearrangement of particles in the first shell causes a small expansion on the surface of the particle cage and generates a harmonic displacement field $u(r)$ in surrounding medium via collective motions of other particles. In bulk systems, one can obtain analytical form of the distortion field by Lifshitz's continuum mechanics analysis \cite{5}
\begin{eqnarray} 
\left(K+\frac{G}{3}\right)\nabla(\nabla.\mathbf{u}) + G\nabla^2\mathbf{u} = 0,
\label{eq:3}
\end{eqnarray}
where $K$ and $G$ are the bulk and shear modulus, respectively. In Ref. \cite{6,7}, the cage expansion amplitude $\Delta r_{eff}$ is found to be
 \begin{eqnarray} 
\Delta r_{eff} = \frac{3}{r_{cage}^3}\left[\frac{r_{cage}^2\Delta r^2}{32} - \frac{r_{cage}\Delta r^3}{192} + \frac{\Delta r^4}{3072} \right].
\end{eqnarray}
Solving Eq. (\ref{eq:3}) with the boundary condition at the cage surface gives
\begin{eqnarray} 
u(r)=\frac{\Delta r_{eff}r_{cage}^2}{r^2}, \quad {r\geq r_{cage}}.
\label{eq:4}
\end{eqnarray}
The spatial harmonic displacement energy of a particle at separation distance $r$ from the center of its arrested cage is $K_0u^2(r)/2$. Since the local time-averaged density is $\rho g(r)$, the number of particles found at a distance between $r$ and $r + dr$ is $\rho g(r)4\pi r^2dr$. One can integrate the elastic energies of particles outside the cage to quantify effects of cooperative motions. The collective elastic barrier, $F_e$, is
\begin{eqnarray} 
F_{e} = 4\pi\rho\int_{r_{cage}}^{\infty}dr r^2 g(r)K_0\frac{u^2(r)}{2}. 
\label{eq:5}
\end{eqnarray}
For $r \geq r_{cage}$, one can approximate $g(r)\approx 1$. The collective motions of molecules play a more important role than the local dynamics in the glass transition at high densities or low temperatures \cite{2,6,7} as depicted in Fig. \ref{fig:1}.

The activated relaxation is governed by both local and non-local processes. If a universal correlation between a local and non-local process is assumed, the total barrier is simply $F_{total} = F_B + F_e$. The alpha relaxation time for a particle to diffuse from its particle cage is quantified by Kramer's theory: 
\begin{eqnarray}
\frac{\tau_\alpha}{\tau_s} = 1+ \frac{2\pi}{\sqrt{K_0K_B}}\frac{k_BT}{d^2} e^{F_{total}/k_BT},
\label{eq:6}
\end{eqnarray}
where $\tau_s$ is a short relaxation time scale. The mathematical form of $\tau_s$ is given by \cite{6,7}
\begin{eqnarray}
\tau_s=g^2(d)\tau_E\left[1+\frac{1}{36\pi\Phi}\int_0^{\infty}dq\frac{q^2(S(q)-1)^2}{S(q)+b(q)} \right],
\label{eq:taus}
\end{eqnarray}
where $\tau_E$ is the Enskog time scale,  $b(q)=1/\left[1-j_0(q)+2j_2(q)\right]$, and $j_n(x)$ indicates the spherical Bessel function of order $n$. Based on many previous studies of thermal liquids, polymers and amorphous drugs \cite{2,6,7,42}, one can assume $\tau_E = 10^{-13}$ s.

To convert our hard-sphere calculations into the temperature dependence of the structural relaxation time, we proposed \cite{42} a thermal mapping, which is based on the thermal expansion process during temperature variation, to convert from a volume fraction of hard-sphere fluid to temperature of experimental material. The mapping is
\begin{eqnarray}
T \approx T_0 - \frac{\Phi - \Phi_0}{\beta\Phi_0}.
\label{eq:7}
\end{eqnarray} 
where $\beta$ is the volume thermal expansion coefficient, $\Phi_0$ and $T_{0}$ are the characteristic volume fraction and temperature, respectively. Since a typical value for linear thermal expansion coefficient of many glass-forming liquids is $2-5\times 10^{-4}$ $K^{-1}$ \cite{31,32}, the volume thermal expansion coefficient $\beta$ is approximately $6-15 \times 10^{-4}$ $K^{-1}$. From a recent work \cite{42}, we estimated $\Phi_0 = 0.50$ and $\beta\Phi_0 = 6\times 10^{-4}$ $K^{-1}$.

The parameter $T_0$ depends on molar mass and particle size. Our numerical calculations indicate the structural relaxation time $\tau_\alpha \approx 100$ s at $\Phi \approx 0.611$. Thus, one can approximately obtain $T_0 = T_g-(0.611-\Phi_0)/\beta\Phi_0$. The experimental value of $T_g$ can be found in many literatures. Based on the calculation, we investigated the temperature dependence of the structural relaxation time of many amorphous drugs and their mixtures (binary and ternary composites) \cite{42}. The theoretical calculations without any adjustable parameters quantitatively agree with various experimental data over 14 decades in time. While simulations can determine relaxation times only over first 3-6 decades and do not access experimental timescale as illustrated in Fig. \ref{fig:1}.

\begin{figure}[htp]
\center
\includegraphics[width=8.5cm]{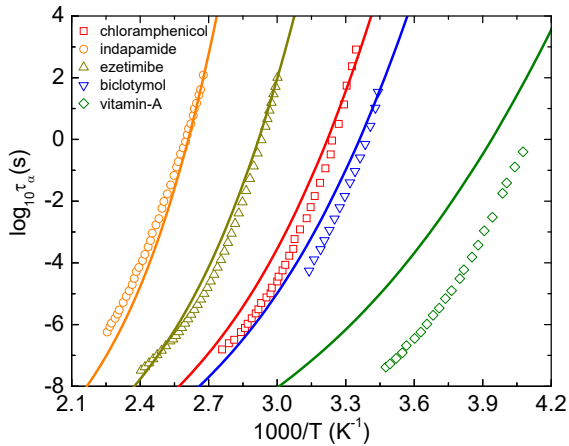}
\caption{\label{fig:2-1}(Color online) The temperature dependence of structural relaxation time of chloramphenicol \cite{38}, indapamide \cite{28}, ezetimibe \cite{28}, biclotymol \cite{40}, and vitamin-A acetate \cite{41}. Open points are experimental data in literatures and solid curves correspond to our ECNLE calculations.}
\end{figure}

Figure \ref{fig:2-1} shows $\log_{10}\tau_\alpha$ of five representative pure amorphous drugs \cite{38,28,40,41} as a function of $1000/T$ calculated using Eqs. (\ref{eq:6}), (\ref{eq:taus}), and (\ref{eq:7}). Overall, the theoretical curves are relatively close to the experimental counterpart, except for calculations of vitamin A. A deviation of experimental data of the vitamin-A drug from theoretical calculations is expected. This is because the theory ignores many chemical and structural complexities such as hydrogen-bonding, network formers, and flexible molecular docking. Moreover, the approach seems to provide less quantitatively accurate predictions for the dynamic fragility of amorphous materials.

\section{Nonuniversal coupling of local and cooperative dynamics}

The dynamic fragility is calculated by
\begin{eqnarray}
m = \left. \frac{\partial\log_{10}(\tau_\alpha)}{\partial(T_g/T)}\right |_{T=T_g}.
\label{eq:fragility}
\end{eqnarray} 
One adopts the physical quantity to classify into two main categories: "strong" or "fragile" for glass-forming materials. For $m \leq 30$, the glass formers are strong. The glass formers having $m \geq 100$ are fragile. The remaining materials are called intermediate glass-forming materials.

The dynamic fragility is very sensitive to the slope of $\tau_\alpha(T)$ at $T = T_g$. Thus, a good agreement between theory and experiment in $\tau_\alpha$ versus $T$ does not mean that another consistency in calculations of $m$ occurs. In Ref. \cite{8}, authors introduced an adjustable parameter $a_c$ to scale the collective elastic barrier as $F_e \rightarrow a_c^2F_e$. The parameter $a_c$ captures chemical and biological complexities, conformational configuration, and chain connectivity in different thermal liquids and polymers. The parameter assesses the relative importance of the collective elastic distortion by assuming a non-universal coupling of the cage-scale hopping and collective rearrangements of fluid particles. The previous work \cite{8} obtained contemporaneously the quantitative accordance between theoretical ECNLE calculations and experimental values of both dynamic fragility and glass transition temperature for 17 polymers. 

\begin{figure}[htp]
\center
\includegraphics[width=8.5cm]{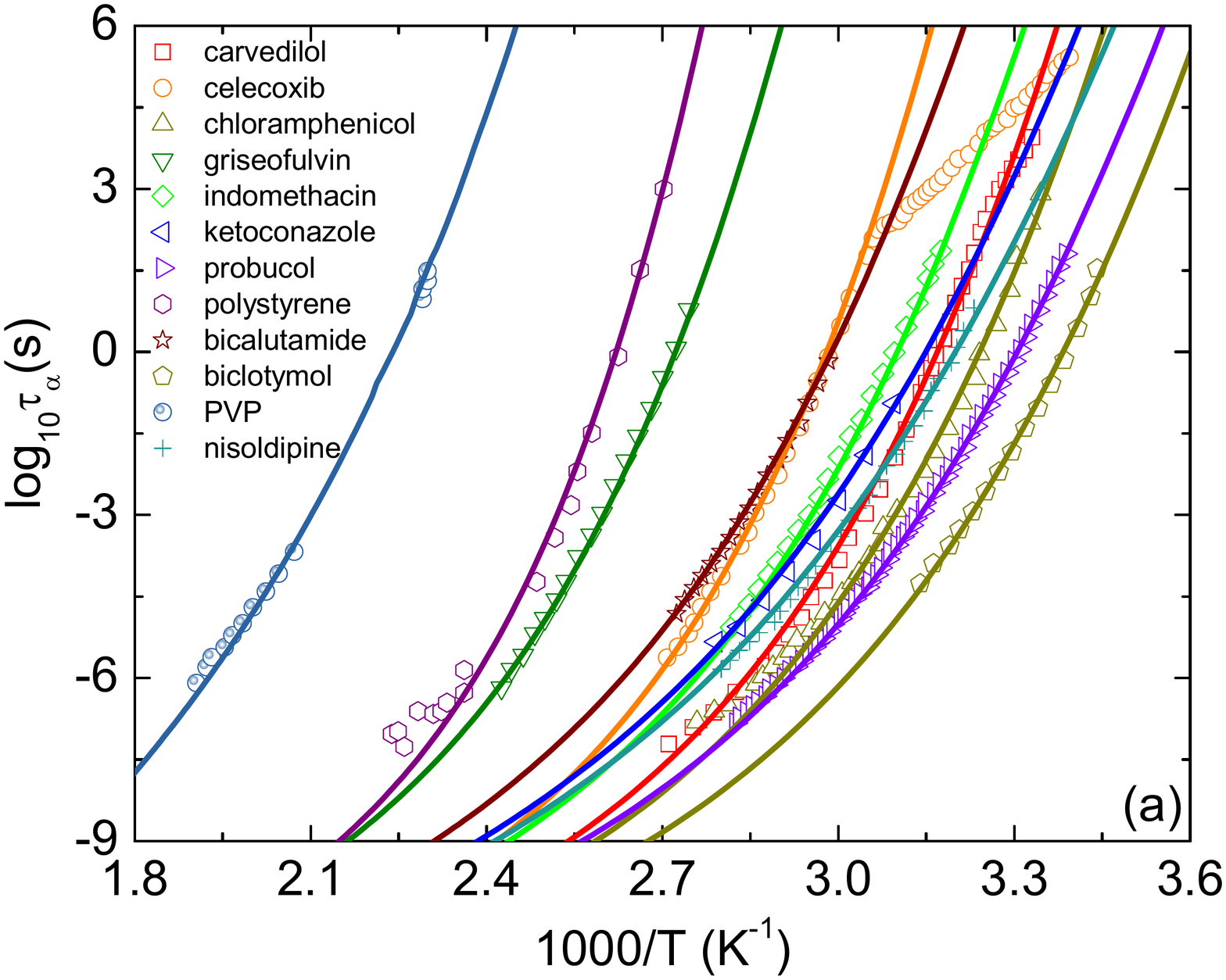}
\includegraphics[width=8.5cm]{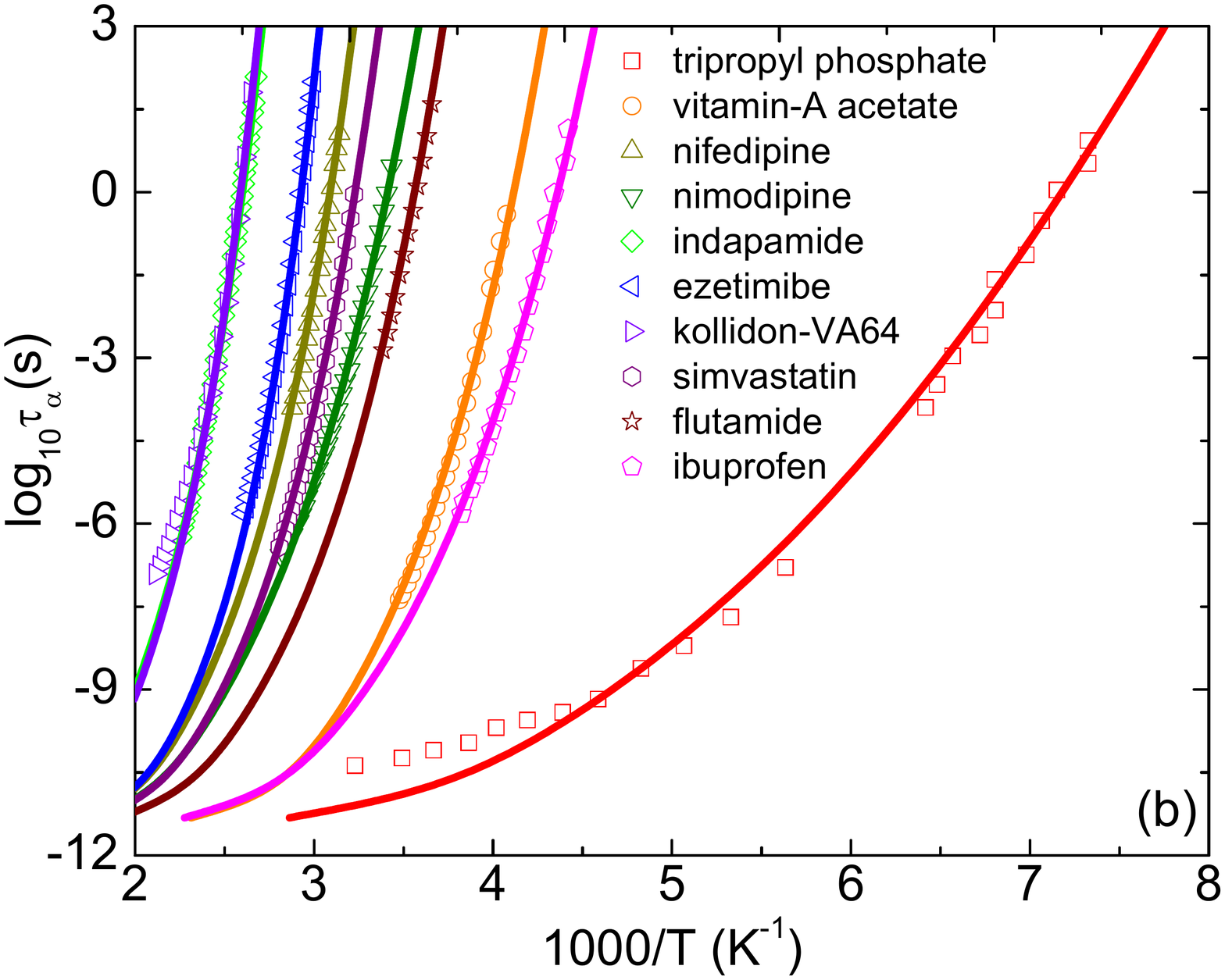}
\caption{\label{fig:2}(Color online) The temperature dependence of structural relaxation time of twenty-two various amorphous drugs and polymers listed in Table \ref{tab:table1} [11, 18-39]. Open points are experimental data in literatures and solid curves correspond to our ECNLE calculations. PVP is an abbreviation of polyvinyl pyrrolidone K30.}
\end{figure}

Motivated by the idea in Ref. \cite{8}, in our calculations, we adjust parameters $T_0$ and $a_c$ to achieve the best fit to the experimental temperature dependence of structural relaxation times. Figure \ref{fig:2} shows experimental data and our theoretical calculations for $\tau_\alpha(T)$ of 22 pure amorphous materials. We carry out the same procedure as calculations in Fig. \ref{fig:2-1} except for now $F_{total} = F_B + a_c^2 F_e$. Our numerical results agree quantitatively well with a wide range of experimental data. Remarkably, the activated events of carvedilol, celecoxib, chloramphenicol, and polystyrene below $T_g$, where $\tau_\alpha$ ranges from 100 s to $10^4$ s, are well-described using the ECNLE theory. Many previous works \cite{29,43,54,55} have observed a distinctive deviation, so-called a dynamic structural decoupling, in the relaxation process at low temperatures. For example, in Fig. \ref{fig:2}a, the growth of $\tau_\alpha(T)$ of carvediol drug below $T_g$ is abruptly deviated from what it is supposed to be. Currently, there is poor theoretical understanding for the interesting but challenging feature. In the framework of the ECNLE theory, the decoupling could be related to a temperature dependence of thermal expansion coefficient $\beta$ in our thermal mapping.  

From calculations in Fig. \ref{fig:2}, we can estimate the glass transition temperature and dynamic fragility. The local-nonlocal coupling parameter $a_c$, the characteristic temperature $T_0$, the melting temperature $T_m$, and theoretical and experimental values for $T_g$ and $m$ of the studied materials are listed in Table \ref{tab:table1}. Clearly, the theoretical $T_g$ is in perfect accordance with the experimental counterpart. The different accuracy of our calculations for the fragility is somehow expected and reflects a complicated $T_g-m$ correlation.

\begin{widetext}
\begin{table*}[htp]
\begin{ruledtabular}
\begin{tabular}{ccccccccc}
\caption{\label{tab:table1}
System parameters and theoretical and experimental results.}
Materials & $T_g$(th) (K) & $T_g$(expt) (K) & $m$(th)& $m$(expt) & $T_m$(expt) (K) & $a_c$  & $T_0$ (K)\\
\hline
carvedilol \cite{33} & 308 & 310 \cite{33} & 91.5 &  &387.5 \cite{19} &2.1 & 450 \\
celecoxib \cite{29,43} & 328 & 328 \cite{29} & 97.8 & 110 \cite{29} & 432 \cite{19} & 2.1& 470 \\
chloramphenicol \cite{38}& 301.1 & 301 \cite{38}& 89 & 116 \cite{38}& 423.5 \cite{19} &2.1 & 443 \\
griseofulvin \cite{34} & 358 & 359 \cite{34} & 88.7 & 84.6 \cite{34} & 489 \cite{18} & 1.2 & 533 \\
indomethacin \cite{36} & 314.1 & 315 \cite{36}& 86.4 & 77, 67, 64 \cite{36}& 432 \cite{18} & 1.5 & 476 \\
ketoconazole \cite{35} & 308 & 316.3 \cite{35}& 71.37 &  & 419 \cite{19} & 1.0 & 493 \\
probucol \cite{39} & 294 & 294.7 \cite{39} & 79.4 &  85 \cite{39} & 398 \cite{19} &1.5 & 456 \\
polystyrene \cite{24} & 373 & 373 \cite{8} & 105.53 & 116, 143, 97, 121 \cite{8}& 513 \cite{20} & 1.7 & 528 \\
bicalutamide \cite{37} & 325.3 & 325.4 \cite{37}& 78.03 & 84 \cite{37}& 464 \cite{18} &1.1 & 505 \\
biclotymol \cite{40}& 288 & 288 \cite{40}& 85.42 & 85 \cite{40}& &2.1 & 430 \\
polyvinylpyrrolidone K30 \cite{29} & 431.2 & 431 \cite{29}& 58.47 & 70 \cite{29}& 523 \cite{22} &0.3 & 672 \\
tripropyl phosphate \cite{30}& 132.1 & 134 \cite{30}& 40.3 &   & 194 \cite{19} &2.4 & 266 \\
vitamin-A acetate \cite{41} & 236.3 & 244.3 \cite{41}& 76.6 & 83 \cite{41} &330 \cite{21} &3.4 & 349 \\
nisoldipine \cite{23}& 303 & 305 \cite{23} & 70.2 & 81 \cite{23} & 425 \cite{21} &1.0 & 488 \\
nifedipine \cite{23}& 315 & 315 \cite{23} & 74.9 & 84 \cite{23} & 444 \cite{18} &1.1 & 494 \\
nimodipine \cite{23}& 284.1 & 285 \cite{23} & 62 & 82 \cite{23} & 398 \cite{19} &0.9 & 474 \\
indapamide \cite{28}& 373.4 & 373.5 \cite{28}& 75.3 & 73 \cite{28}& 433 \cite{19}& 0.7 & 578 \\
ezetimibe \cite{28} & 333.3 & 333.1 \cite{28} & 91.1 & 93 \cite{28} & 436 \cite{19} & 1.5 & 495 \\
kollidon VA64 \cite{42} & 376.2 & 378 \cite{42} & 79 &   & & 0.8 & 573 \\
simvastatin \cite{42}& 301 & 303 \cite{42} & 74.8 & 73 & 411 \cite{19} &1.3 & 471 \\
flutamide \cite{26} & 272 & 271 \cite{26}& 71.6 &   & 385 \cite{19} &1.4 & 438 \\
ibuprofen \cite{25} & 222 & 225 \cite{25} & 68 & 87 \cite{25} & 346 \cite{18} &2.4 & 356 \\
\end{tabular}
\end{ruledtabular}
\end{table*}
\end{widetext}

\section{Machine-Learning based analysis of glassy dynamics}
There are two main methods to obtain $T_g$ from experiments. First, one can use the DSC measurement for samples to find the value of $T_g$. Second, experimentalists have widely used the Vogel-Fulcher-Tammann (VFT) equation to fit data measured by the BDS, which is inaccessible to the deeply supercooled state or ($\tau_\alpha \leq 1 s $). From this fitting function, they can obtain the glass transition temperature at $\tau_\alpha = 100$ $s$ via extrapolation. However, the fitting depends strongly on a focused regime of data \cite{14}. The selection process possibly causes large overprediction or underprediction of $T_g$. This is possibility why many works have reported different $T_g$ values for a given material.

\begin{figure}[htp]
\center
\includegraphics[width=8.5cm]{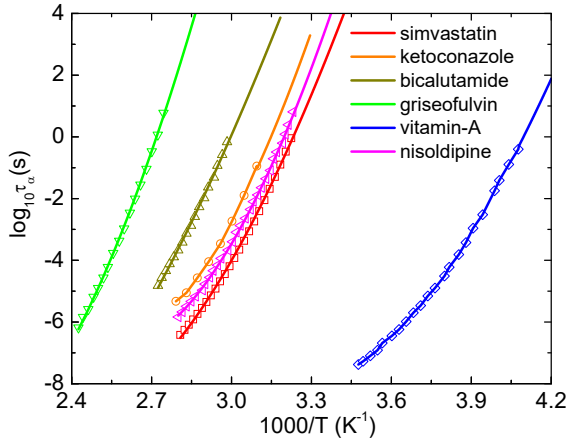}
\caption{\label{fig:3}(Color online) The temperature dependence of structural relaxation time of simvastatin \cite{42}, ketoconazole \cite{35}, bicalutamide \cite{37}, griseofulvin \cite{34}, vitamin-A acetate \cite{41}, and nisoldipine \cite{23}. Open points are experimental data in literatures and solid curves correspond to machine learning-based calculations.}
\end{figure}

Here we introduce, for the first time, another approach based on machine learning techniques. After obtaining a predictive model by applying the support vector regression (SVR) in the scikit-learn Python library \cite{60} to experimental data of the temperature dependence of $\tau_\alpha$ at high temperatures, we can predict new relaxation times at lower temperatures than the coolest temperature in the training dataset. The SVR with the radial basis function (RBF) kernel has two controlled parameters including a regularization parameter $C_{RBF}$ and a RBF kernel parameter $\gamma$. While $\gamma$ is considered as the inverse of the standard deviation of the RBF kernel, $C_{RBF}$ determines the penalty of large slack variables. In our calculations, we chose $\gamma = 0.1$ and obtained equal performance when $C_{RBF} \geq 100$. As shown in Fig. \ref{fig:3}, $\tau_\alpha(T)$ versus $1000/T$ predicted by the SVR with $C_{RBF}=10^3$ grows smoothly and are close to experimental data for some representative amorphous drugs. The number of BDS data points for simvastatin, ketoconazole, bicalutamide, griseofulvin, vitamin-A acetate, and nisoldipine used for training are 24, 8, 17, 17, 22, and 28, respectively. The SVR calculations give $T_g$s for simvastatin, ketoconazole, bicalutamide, griseofulvin, vitamin-A acetate, and nisoldipine are 300.2, 308.64, 323, 358.2, 237.8, and 304 K, respectively. These values quantitatively agree experimental data and our ECNLE results in Table \ref{tab:table1}. This is a new reliable approach for investigating molecular dynamics of amorphous materials near $T_g$, particularly when the structural decoupling appears \cite{29,43,37,54,55}. The linear regression cannot be used in this protocol since it enforces a Arrhenius nature on any experimental data and keeps it unchanged in the predictive process.

\begin{figure}[htp]
\center
\includegraphics[width=8.5cm]{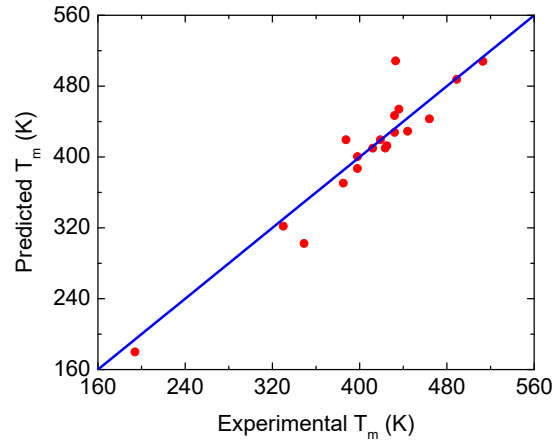}
\caption{\label{fig:4}(Color online) Cross-plot of predicted and experimental values of the melting temperature for all 20 drugs and polymers from Table \ref{tab:table1}. The blue line indicates perfect agreement.}
\end{figure}

To find new minimalist correlations or physical insights among the quantities in Table \ref{tab:table1}, we employ a linear regression model in the scikit-learn library \cite{60}. This regression algorithm provides the simplest relation to describe a target variable from a set of descriptor variables. By adopting $T_g$ and $T_m$ experimental values of 71 glassy drugs listed in Ref. \cite{15} as a training dataset, one obtains $T_m \approx 1.362 T_g$. It is important to note that these drugs are different from amorphous materials in this work. Then, we apply the predicted relation to our twenty substances to evaluate the validity of the model. Numerical results in Fig. \ref{fig:4} indicate that the $T_g-T_m$ correlation works well. This finding suggests the ECNLE theory can be exploited to estimate the melting temperature with a reasonable deviation. 

In many prior works \cite{46,47,48}, the melting point of amorphous drugs exhibits an essential role in the solubility determination of the drugs. The linear relation between $T_g$ and $T_m$ suggests that one can employ ECNLE theory to evaluate the solubility of amorphous materials and their mixtures.

\section{effects of cooling rate on glassy dynamics}

Although it is experimentally well-known that cooling rate has considerable impact on glassy dynamics, but theoretical understanding has remained ambiguous. Thus, in this section, we would propose a simple model to estimate how $T_g$ is varied with different cooling rates. 

According to an assumption introduced by Cooper and Gupta in 1982 \cite{44}, the molecular relaxation time is approximately equal to the experimental observation time at $T_g$. This assumption leads to  a new definition of a cooling rate, $h$,
\begin{eqnarray}
h = -\frac{dT}{dt} \approx -\frac{dT}{d\tau_\alpha}.
\label{eq:8}
\end{eqnarray}
The minus sign in Eq.(\ref{eq:8}) represents an inverse variation between mobility and temperature. 

Near glass transition temperature, $e^{F_{total}/k_BT} \gg 1$ and the alpha structural relaxation time in Eq. (\ref{eq:6}) approximately becomes
\begin{eqnarray}
\tau_\alpha(T) \approx \frac{2\pi\tau_s}{\sqrt{K_0K_B}}\frac{k_BT}{d^2}e^{F_{total}/k_BT}.
\label{eq:9}
\end{eqnarray}
Recall that the total barrier here is $F_{total} = F_B + a_c^2 F_e$ as discussed in Section III. After straightforward transformations, one obtains

\begin{eqnarray}
\frac{1}{\tau_\alpha}\frac{d \tau_\alpha}{d T} &=& \frac{d}{dT}\ln\left( \frac{2\pi\tau_s}{\sqrt{K_0K_B}}\frac{k_BT}{d^2}\right) +\frac{d}{dT}\left( \frac{F_{total}}{k_BT}\right).\nonumber\\
\label{eq:9-1}
\end{eqnarray}
Since $\tau_s$ is order of picoseconds ($10^{-12}$ s), the first term is much smaller than the second term near $T_g$. Thus,  
\begin{eqnarray}
\left.h\tau_\alpha(T_g)\frac{d}{dT}\left( \frac{F_{total}}{k_BT}\right)\right|_{T=T_g} \approx -1.
\label{eq:10}
\end{eqnarray}
When the temperature dependence of structural relaxation time obeys the Arrhenius behavior, the total barrier $F_{total}$ is a constant and our Eq. (\ref{eq:10}) can be deduced to be the same mathematical form as previous studies \cite{12,13}
\begin{eqnarray}
h\tau_\alpha(T_g)\frac{F_{total}}{k_BT_g^2}=1 \quad \ce{or} \quad \frac{h\tau_\alpha(T_g)}{T_g^2}   = \ce{constant} .
\label{eq:11}
\end{eqnarray}

Moreover, combining Eq. (\ref{eq:9}) and (\ref{eq:10}) gives
\begin{eqnarray}
&&\ln\left(-\left.\frac{d}{dT}\left( \frac{F_{total}}{k_BT}\right)\right|_{T=T_g}\right)+  \ln\left(\frac{2\pi \tau_s}{\sqrt{K_0K_B}}\frac{k_BT_g}{d^2} \right) \nonumber\\ &+& \frac{F_{total}}{k_BT_g} +\ln h = 0. 
\label{eq:15}
\end{eqnarray}

The above equation reveals explicitly a correlation between the cooling rate and the glass transition temperature.  Recall that various experimental studies have empirically indicated that $\ln h$ is linearly proportional to  $-1/T_g$ in a specific range of $T_g$. If the relation is universal, it suggests the first two terms in Eq.(\ref{eq:15}) may play a minor role compared others and then our analysis can clearly explain the experimental observation.

\begin{figure}[htp]
\center
\includegraphics[width=8.5cm]{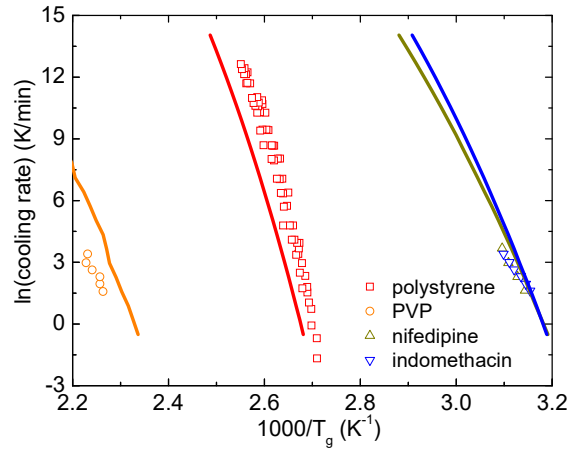}
\caption{\label{fig:5}(Color online) The logarithm of cooling rate as a function of inverse glass transition temperatures of various amorphous drugs and polymers. Points are experimental data and solid curves correspond to our theoretical calculations using Eq.(\ref{eq:10}).}
\end{figure}

Figure \ref{fig:5} shows how the cooling rate influences the glass transition temperature of polystyrene \cite{17}, PVP \cite{11}, nifedipine \cite{16}, and indomethacin \cite{11}. Our theoretical calculations are relatively consistent with experimental data. Although both theoretical curves and data points are not perfect straight lines, the rough linearity is possibly observed in a certain range of the glass transition temperature. One can also crudely view these smooth curves as a combination of high- and low-$T_g$ linear branches. Furthermore, as discussed in Section IV, $T_m \approx 1.362 T_g$, we expect plotting $\ln h$ versus $1000/T_m$ (not shown here) does not change the variation trends compared to Fig. \ref{fig:5}. The finding is in accordance with other experimental works \cite{45}. 

The effects of cooling rate on $T_g$ can be also analyzed using the VFT-type relaxation time, which is
\begin{eqnarray}
\tau_\alpha(T) = \tau_0\exp\left(\frac{DT_{VFT}}{T-T_{VFT}} \right),
\label{eq:12}
\end{eqnarray}
where $\tau_0$, $D$, and $T_{VFT}$ are parameters fitted from experimental data. At $T = T_g$, we have
\begin{eqnarray}
h\tau_\alpha(T_g) = \frac{(T_g-T_{VFT})^2}{DT_{VFT}}.
\label{eq:13}
\end{eqnarray}

Interestingly, both ECNLE theory and VFT-function analysis give us the same form of the nontrivial correlation among the cooling rate, glass transition temperature, and fragility 
\begin{eqnarray}
h\tau_\alpha(T_g) = \frac{T_g}{m\ln(10)}.
\label{eq:18}
\end{eqnarray}
For small $\ln q$ ($\leq 0$) or low cooling rates, $T_g$ of polystyrene is nearly unchanged, approximately 373 $K$. Thus, one can utilize the definition of $\tau_\alpha(T_g) = 100$ s to estimate $T_g$ in the range of the cooling rate. It suggests that if $\tau_\alpha$ of all substances is measured at the same low cooling rate, $h\tau_\alpha(T_g)\approx 100h$ and $T_g$ is linearly proportional to $m$. However, since different experiments have been carried out at different cooling rates, it is hard to obtain a $T_g-m$ trivial correlation.
\section{Conclusions}
We have shown several theoretical approaches to improve quantitatively accurate predictions of the glass transition temperature and dynamic fragility of twenty-two amorphous drugs and polymers. The temperature dependence of the structural relaxation time is theoretically calculated using the version of ECNLE theory. By introducing an adjustable parameter to describe a non-universal correlation between local and collective molecular dynamics in different materials, our numerical results for the dynamic fragility and $T_g$ measured at various cooling rates show better quantitative agreement with experiments than simply using the universal local-nonlocal coupling. Applying machine-learning calculations to BDS experimental data gives the same $T_g$ values as using the VFT fit function. The finding suggests that machine learning technique can verify the VFT-based results in all BDS studies instead of comparing with DSC experiments. The machine-learning calculation may be more reliable to predict $T_g$ and $m$ when the structural decoupling of relaxation process occurs.  Machine learning also reveals the linear relation of $T_m \approx 1.362 T_g$. This relation explains why the cooling rate changes the melting point and glass transition temperature in the same manner.

\section*{Conflicts of interest}
There are no conflicts to declare.

\begin{acknowledgments}
This work was supported by JSPS KAKENHI Grant Numbers JP19F18322 and JP18H01154. The author, M.P., is grateful for the financial support received within the Project no. 2015/16/W/NZ7/00404 (SYMFONIA 3) from the National Science Centre, Poland. This research was funded by the Vietnam National
Foundation for Science and Technology Development (NAFOSTED) under grant number 103.01-2019.318.
\end{acknowledgments}

\end{document}